\documentclass[sigplan,nonacm]{acmart}

\usepackage{tabularx}
\usepackage{enumitem}

\begin{document}

\title{PATTON: Enabling Commodity PIM for Production LLM Serving}

\author{Hangyeol Kim}
\email{khk070623@kaist.ac.kr}
\affiliation{%
  \institution{KAIST}
  \city{Daejeon}
  \country{Republic of Korea}
}

\author{Sanghyun Lee}
\email{lysanghyun@kaist.ac.kr}
\affiliation{%
  \institution{KAIST}
  \city{Daejeon}
  \country{Republic of Korea}
}

\author{Teokkyu Suh}
\email{ejrrb102@kaist.ac.kr}
\affiliation{%
  \institution{KAIST}
  \city{Daejeon}
  \country{Republic of Korea}
}

\author{Joo-Young Kim}
\email{jooyoung1203@kaist.ac.kr}
\affiliation{%
  \institution{KAIST}
  \city{Daejeon}
  \country{Republic of Korea}
}

\begin{abstract}
  Processing-in-Memory (PIM) is a promising architecture for accelerating memory-bound decode stage attention. However, attention acceleration alone is insufficient for production LLM serving, where serving engines dynamically allocate, populate, share, cache, and reclaim logical KV cache blocks. Supporting this block lifecycle on commodity PIM requires allocating physical memory, mapping logical block IDs and token offsets to exact physical addresses, and generating corresponding command sequences. Block-to-PIM mapping determines GEMV and KV cache write efficiency, while memory allocation granularity determines capacity efficiency. Our observations reveal a fundamental conflict among these three objectives for the Value cache: a GEMV-optimized layout scatters each newly generated Value vector across multiple rows, making decode stage single token Value writes a substantial component of attention latency. Allocating an entire row to each request preserves long GEMV reductions but wastes capacity, whereas fine-grained sharing improves capacity efficiency but fragments those reductions. We present \textit{PATTON}, a PIM runtime system that integrates production LLM serving engines with commodity PIM. PATTON introduces \textit{hierarchical granule allocation}: block-sized Key and Value granules map one-to-one to logical token blocks, fixing their physical placements and commands, while coarser granules group them to improve GEMV execution and capacity efficiency. Its \textit{Commit Zone} stages partial Value blocks for efficient single token writes before committing them to GEMV-optimized locations. PATTON records these placements to generate KV cache writes and commands for QK$^\top$ and SV. Across attention execution and runtime-induced prefill recomputation, PATTON achieves an average $1.95\times$ speedup and $4.83\times$ higher energy efficiency over the evaluated baselines. PATTON requires no modifications to the PIM processing units and maintains a KV cache hit rate comparable to vLLM with its native GPU KV cache.
\end{abstract}

\maketitle

\section{Introduction}
Large language models (LLMs) are increasingly deployed through production serving engines such as vLLM, SGLang, and TensorRT-LLM~\cite{vllm,sglang,tensorrtllm}. Unlike standalone inference, these engines continuously process dynamically arriving requests using scheduling techniques such as continuous batching~\cite{orca,vllm,sarathi,sarathi-serve}.
They also manage the key-value (KV) cache using block-based mechanisms such as PagedAttention, which enables prefix sharing, prefix caching, and efficient dynamic memory management~\cite{vllm,cachedattention}. Together, these techniques create a dynamic serving workflow in which the engine continuously allocates, populates, shares, caches, and reclaims KV cache blocks as requests progress. Therefore, a hardware platform for production LLM serving must efficiently support these KV cache operations in addition to accelerating attention computation.

Meanwhile, attention remains fundamentally memory-bound~\cite{flat,dao2022flashattention,dao2024flashattention,gholami2024ai}, and its arithmetic intensity increases only marginally as batch size grows~\cite{attacc}. This property has motivated Processing-in-Memory (PIM)~\cite{ahn2015scalable,ahn2015pim,asghari2016chameleon,cho2021accelerating,choi2023unleashing,gao2019computedram,giannoula2022sparsep,gomez2022machine,gomez2022benchmarking,he2020newton,hyun2024pathfinding,kim2021aquabolt,kwon2019tensordimm,laguna2021memory,lee2021hardware,oliveira2022accelerating,park2021trim,park2024lpddr,aim,cent}, whose high internal memory bandwidth is well suited to memory-bound attention~\cite{attacc,attenpim,blockpim,neupims,pimphony,l3,cent,pimmalloc,li2024specpim,paise}. Prior PIM-based LLM systems exploit this advantage through heterogeneous xPU+PIM architectures~\cite{neupims,attacc,pimphony,blockpim,l3,he2025papi,paise}: the xPU executes compute-bound operations such as feedforward network (FFN) layers, while PIM executes decode stage attention.

\textit{However, enabling commodity PIM for production LLM serving requires a PIM runtime that speaks the languages of both the serving engine and PIM---a capability that remains underexplored.} Serving engines express KV cache operations using logical blocks, which are software-visible blocks identified by block IDs rather than physical memory addresses. To preserve these semantics, a PIM runtime must allocate physical memory and map each logical block and token offset to an exact channel-bank-row-column address. It must then translate each operation into the corresponding PIM command sequence. Efficiently realizing this translation is challenging because commodity PIM follows a rigid execution model in which computation patterns are constrained by the physical organization of data in memory. The block-to-PIM mapping determines both the DRAM write patterns of newly generated Key and Value vectors and the scheduling of multiply-accumulate (MAC) operations for attention execution. Also, how finely memory is allocated around these mapped blocks determines how efficiently the PIM memory space is utilized.

Existing PIM systems address only parts of this translation problem. Some represent the KV cache as a large tensor and map it to a large PIM memory region at initialization~\cite{cent,attacc,paise}, limiting support for dynamic, block-level KV cache management. Systems with finer-grained memory allocation~\cite{neupims,pimphony,pimmalloc,blockpim} dynamically provision physical memory but do not specify how logical token blocks are physically organized within the allocated space. Other systems optimize DRAM writes and GEMV execution but require significant modifications to the processing units~\cite{l3,attenpim}. These limitations collectively leave three key challenges unresolved.

\begin{itemize}[leftmargin=*, labelsep=0.5em]
    \item \textbf{The runtime must bridge the gap between the serving engine's logical block management and PIM's physical memory organization:} The serving engine specifies KV cache writes and attention operations using logical block IDs and token offsets. For a KV cache write, the runtime must translate these into the physical address of the corresponding Key or Value vector. For attention execution, it must identify the physical regions of the blocks participating in Query-Key multiplication (QK$^\top$) and Score-Value multiplication (SV) and generate their MAC command sequences. Supporting both operations requires an explicit mapping from each logical block and token offset to its channel-bank-row-column placement in PIM.

    \item \textbf{Block-to-PIM mapping and memory allocation must jointly account for Key and Value characteristics, GEMV efficiency, KV cache write efficiency, and capacity efficiency:} QK$^\top$ reduces over the fixed head dimension, whereas SV reduces over the growing token dimension, requiring different physical block compositions for Key and Value. GEMV efficiency favors long contiguous reductions executed in parallel across channels, while KV cache write efficiency favors high burst utilization and few row activations. Memory allocation must be fine-grained enough to use PIM capacity efficiently, thereby retaining useful cached prefixes and preserving the KV cache hit rate. The runtime must therefore coordinate block-to-PIM mapping with memory allocation to satisfy all three efficiency objectives.

    \item \textbf{This joint objective exposes a fundamental conflict in Value layout:} A GEMV-optimized Value layout places consecutive tokens along the columns of each row to accumulate them in one continuous reduction. However, the same layout scatters a newly generated token's Value vector across multiple rows, causing poorly utilized row activations and substantial write latency. Allocating an entire row to each request preserves GEMV efficiency but wastes capacity, potentially forcing the eviction of useful cached prefixes. Naively allowing multiple requests to share a row recovers that capacity but can fragment each request's reduction. The runtime must therefore handle Value placement with particular care.
\end{itemize}

\begin{figure}[t]
    \centering
    \includegraphics[width=\columnwidth]{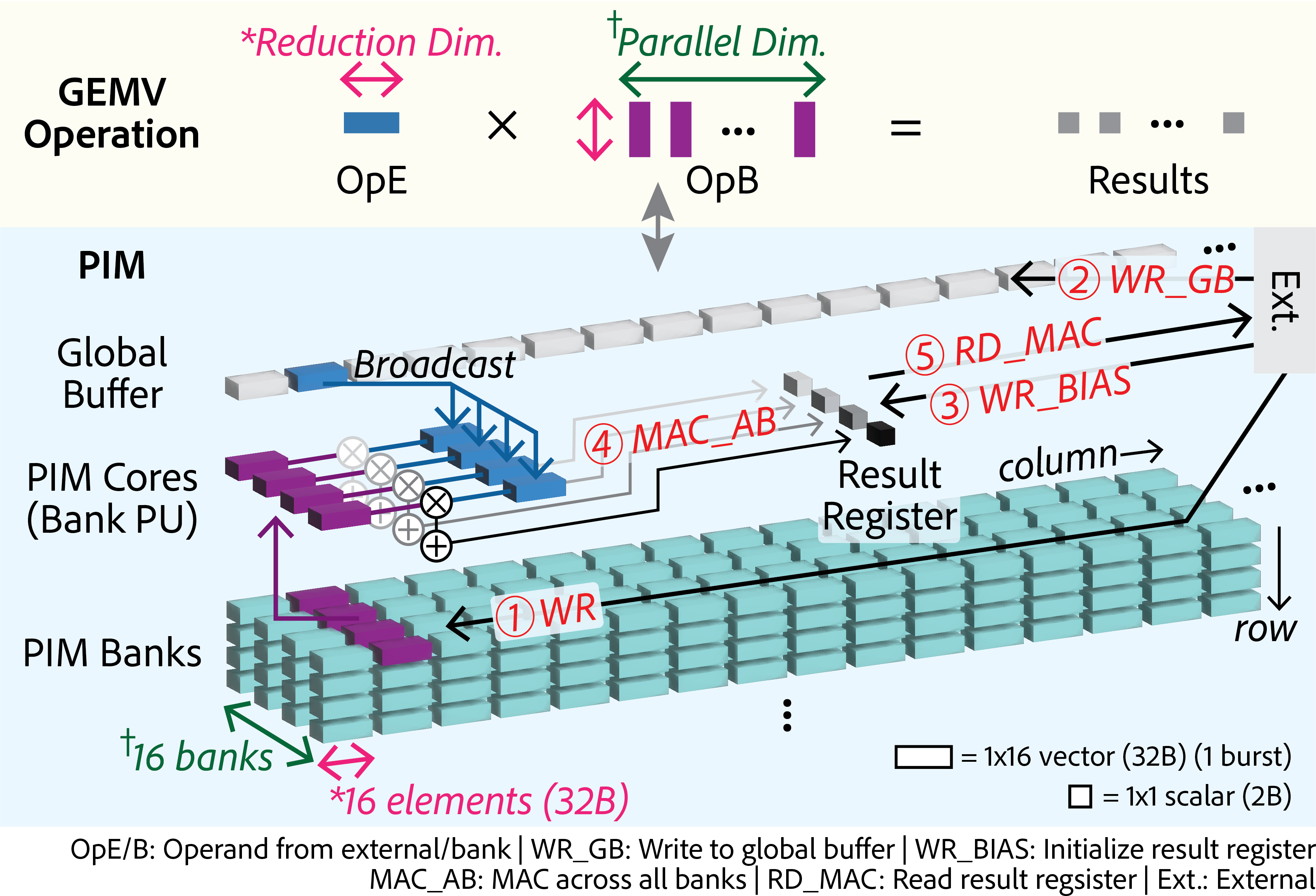}
    \caption{Commodity PIM architecture and the mapping of a GEMV operation to PIM.}
    \label{fig:background1}
\end{figure}

\begin{figure}[t]
    \centering
    \includegraphics[width=\columnwidth]{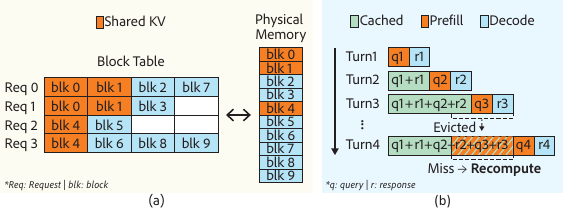}
    \caption{(a) Block-level KV cache management and prefix sharing. (b) Prefix caching in multi-turn conversation scenarios.}
    \label{fig:background2}
\end{figure}

To address these challenges, we propose \textbf{PATTON}, a runtime system for production LLM serving on commodity PIM architectures. PATTON performs block-to-PIM translation through \textbf{hierarchical granule allocation} and a \textbf{Commit Zone}. Hierarchical granule allocation composes minimum valid parallel PIM execution granules into block-sized Key and Value granules, each mapped one-to-one to a logical token block to fix its placement and commands. Coarser granules group these block mappings across channels, rows, and columns to exploit channel parallelism, optimize GEMV execution, and use PIM capacity efficiently. For Value, this grouping keeps several neighboring block locations available to one request so that SV retains a useful continuous reduction, while allowing other requests to use the remaining row capacity. The Commit Zone temporarily stores the Value block currently being filled in a layout optimized for decode stage single token writes. After all token positions have been written, PATTON \textit{commits} the block to its reserved GEMV-optimized location. PATTON records these bindings to translate logical block IDs and token offsets into KV cache DRAM writes and command sequences for QK$^\top$ and SV.

We implement PATTON in vLLM, a representative production serving engine, and target the AiM architecture, whose instruction set is publicly available~\cite{aim}. Across attention execution and runtime-induced prefill recomputation, PATTON achieves an average $1.95\times$ speedup and $4.83\times$ higher energy efficiency over the evaluated baselines without modifying the underlying PIM architecture.

\section{Background}
\label{sec:background}

\subsection{GEMV Execution on Commodity PIM}

Commodity PIM devices accelerate GEMV using a MAC processing unit (PU) in each bank, while the banks in a channel operate in parallel using a shared input operand. Accelerator-in-Memory (AiM) is a GDDR6-based PIM device representative of this commodity organization, and its instruction set is publicly available~\cite{aim}.

Fig.~\ref{fig:background1} shows the AiM architecture and GEMV execution. The target configuration contains 32 channels, each with one Global Buffer (GB) shared by 16 banks. Each bank contains one MAC PU, and the GB supplies the same input operand to all 16 PUs, allowing the banks to compute 16 matrix outputs in parallel. GEMV data placement follows this organization: the parallel dimension is distributed across banks, and the reduction dimension is placed along consecutive columns within each bank row. The input vector is loaded into the GB. After the target rows are activated, one MACAB command selects a 32\,B burst containing 16 16-bit matrix elements from every bank and applies the corresponding 16 16-bit input elements from the GB. Across the 16 banks, the command computes $(1\times16)\times(16\times16)$ and produces 16 MAC results in parallel. The MAC results are retained in per-bank result registers, allowing subsequent MACAB commands to continue the reduction over additional columns. A 2\,KB bank row contains 64 burst positions, so one row activation can support up to 64 consecutive MACAB commands. After the reduction completes, RDMAC reads the 16 result registers in parallel and returns 32\,B per channel.

\subsection{AiM Command Set}

AiM extends the standard GDDR6 interface with commands for deep learning operations~\cite{aim}. WRGB loads an input vector into the GB, and WRBIAS initializes the per-bank result registers. MACAB invokes the MAC PUs in all banks, while RDMAC reads the result registers. RDAF reads the outputs produced by the activation-function unit. WRGB and WRBIAS transfer input data into the PIM datapath, while RDMAC transfers MAC results out. Repeated input loading, register initialization, and result retrieval incur PIM I/O overhead and must be minimized for efficient GEMV execution~\cite{pimphony}. A data layout should therefore reuse GB operands and continue a reduction across as many MACAB commands as possible before retrieving the results. AiM also provides commands for internal data movement. RDCP copies data from a bank to the GB, and WRCP copies data from the GB to a bank. These commands prepare operands and retrieve results without changing the MAC datapath.

\subsection{PagedAttention and Prefix Sharing}

A request's final context length is not known in advance. Static KV cache allocation reserves contiguous memory for the maximum supported length of every request. Most requests do not use this entire region, wasting memory and limiting the batch size. vLLM addresses this problem with PagedAttention, which applies OS paging principles to the KV cache~\cite{vllm,vllm-github}. The KV cache is divided into fixed-size logical blocks, typically containing 16 tokens. Physical blocks are allocated on demand as tokens are appended. As shown in Fig.~\ref{fig:background2}(a), each request has a block table that maps its logical blocks to noncontiguous physical blocks in memory. The attention kernel uses this table to access the request's KV cache. This design avoids reserving memory for the maximum context length and limits unused space to the final partially filled block of each request. Block-based management also enables prefix sharing~\cite{vllm,vllm-github}. Requests with the same prefix can reference the same full physical blocks, avoiding redundant KV cache storage and computation. A shared block is read-only. The final, partially filled block remains private so that new tokens can be appended.

\begin{figure*}[t]
    \centering
    \includegraphics[width=\textwidth]{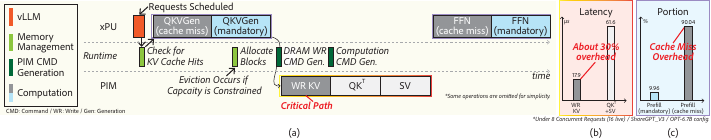}
    \caption{(a) The PIM runtime works with vLLM to manage PIM memory and generate commands for KV cache writes, QK$^\top$, and SV. (b) KV cache writes add about 30\% of QK$^\top$ and SV latency. (c) Without prefix caching, cache misses account for about 90\% of prefill work.}
    \label{fig:wrkvoverhead}
\end{figure*}





\subsection{Multi-Turn LLM Serving}
\label{sec:multi-turn-serving}

Multi-turn conversation is a common interaction pattern in LLM applications, where a user and an LLM iteratively exchange queries and responses. Each query-response interaction constitutes a single LLM inference request, referred to as a turn. As shown in Fig.~\ref{fig:background2}(b), let $q_i$ and $r_i$ denote the user query and LLM response at the $i$-th turn, respectively. At turn $N+1$, the LLM processes $q_1r_1q_2r_2\ldots q_Nr_Nq_{N+1}$ to generate $r_{N+1}$. Consequently, multi-turn conversations naturally exhibit strong prefix reuse, as the context from previous turns becomes a prefix of each subsequent request.

Prefix caching exploits this reuse opportunity~\cite{cachedattention}. As illustrated in Fig.~\ref{fig:background2}(b), vLLM retains a completed request's KV cache blocks as zero-reference cached blocks, allowing later requests to reuse matching prefixes without recomputation~\cite{vllm,vllm-github}. These blocks remain cached until capacity pressure forces their eviction. If a later turn needs an evicted block, that portion of the prefix becomes a cache miss and must be recomputed. Thus, retaining cached prefixes across inter-turn intervals is critical to avoiding redundant recomputation in multi-turn LLM serving.

\section{Motivation and Observations}
\label{sec:motivation}

\subsection{Runtime Responsibilities in Production Serving}

Production LLM serving engines such as vLLM manage the KV cache as logical token blocks and allocate blocks on demand as requests progress~\cite{vllm,vllm-github,cachedattention}. vLLM expresses KV cache operations using logical block IDs and token offsets while remaining agnostic to the physical PIM address space. Supporting this interface requires the PIM runtime to jointly determine each block's physical placement and allocate PIM memory at a granularity that preserves capacity efficiency. The runtime retains the resulting metadata for cache-hit checks, KV cache writes, and attention execution.

Fig.~\ref{fig:wrkvoverhead}(a) shows how this metadata is created and used. After scheduling requests, vLLM identifies which cached logical blocks remain resident and which have been evicted. Inefficient use of PIM capacity can evict cached KV blocks that would remain resident in a conventional GPU KV cache with the same nominal capacity, causing additional cache misses and recomputation. The scheduled batch therefore contains both tokens from cache-missed blocks that require recomputation and tokens that must be computed regardless of cache reuse.

The xPU generates Query, Key, and Value vectors for the scheduled batch. For newly generated or recomputed blocks, the runtime allocates PIM memory for their Key and Value data and records the assigned locations as placement metadata. Using this metadata, it generates the DRAM commands that write each Key and Value vector to its assigned location. The runtime also uses the metadata to identify the physical regions for QK$^\top$ and SV, form the corresponding GB operands, issue MAC commands, and retrieve and reinitialize the accumulators at the appropriate boundaries. PIM performs the KV cache writes and attention operations, after which the xPU performs the FFN computation.

\begin{figure*}[t]
    \centering
    \includegraphics[width=\textwidth]{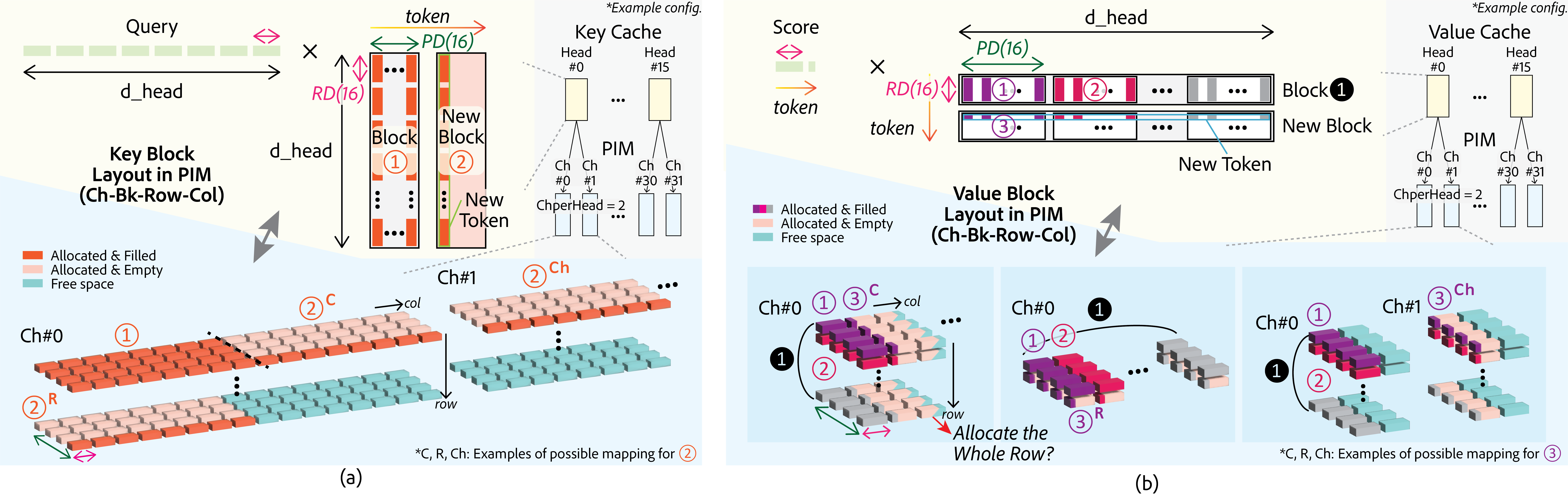}
    \caption{Column (Case C), row (Case R), and channel (Case Ch) placement choices. (a) For Key, Case Ch$\rightarrow$C combines channel parallelism with compact placement. (b) For Value, Case Ch$\rightarrow$C preserves channel parallelism and contiguous SV reductions, whereas Case R reduces row activations for Value writes. Whole-row reservation preserves the reduction but may leave capacity unused.}
    \label{fig:motivationtop}
\end{figure*}

\subsection{Key and Value Placement Tradeoffs}

Designing the physical layout required by this execution path demands separate consideration of the Key and Value caches rather than one uniform mapping for both. As described in Section~\ref{sec:background}, PIM maps a GEMV reduction dimension along columns within a row and distributes the parallel dimension across banks. In QK$^\top$, each token produces an independent output and the reduction dimension is the fixed head dimension, $d_{\mathrm{head}}$. In SV, each element of the head dimension is an independent output and the reduction dimension is the token dimension, which grows with the context.

\subsubsection{Key cache placement}

For Key, the 16 tokens in a block naturally occupy the 16 banks, while the $d_{\mathrm{head}}$ elements of each token span consecutive columns. Once these elements have been processed, the dot products for that block are complete; the next Key block represents independent outputs and cannot extend the same reduction. Whether one block can utilize a complete bank row therefore depends on the head dimension.

A head dimension of 128 is widely used across LLMs. We therefore use it as the representative configuration for explanation, while PATTON does not require this specific value. With BF16 elements, one Key vector occupies only 256\,B, one eighth of a 2\,KB bank row. Thus, one Key block uses only a fraction of each bank row.

Fig.~\ref{fig:motivationtop}(a) shows three placements for a new Key block. Case C places it in unused burst columns of the same row, compacting the physical footprint and avoiding an additional row. Case R places it at the same burst offsets in another row, allowing the Query operand to reuse the same GB offsets. Unless later blocks fill its unused columns, however, Case R leaves seven eighths of the row empty. Case Ch places the block in another channel assigned to the same KV head, allowing QK$^\top$ operations to exploit channel-level parallelism. We use \emph{ChPerHead} to denote the number of channels assigned to each KV head.

Based on this observation, PATTON applies Case Ch first and Case C second: it distributes blocks across ChPerHead channels and then fills unused columns in their rows. This Case Ch$\rightarrow$C order combines channel parallelism with compact row use. Each Key block still retains a separate WRBIAS and RDMAC boundary because its output is independent.

\subsubsection{Value cache placement}

Value exposes a more fundamental conflict because its token dimension is the reduction dimension. A Value block therefore provides exactly 16 reduction elements for each output, matching one burst. The head dimension forms the parallel dimension; with a head dimension of 128 and 16 banks, one Value block is divided into eight parallel subblocks. Fig.~\ref{fig:motivationtop}(b) shows how Cases C, R, and Ch compose these subblocks and consecutive Value blocks.

Case C places the corresponding subblock of the next Value block in the next burst column of the same row, extending the token reduction. Case Ch places work in another of the ChPerHead channels and provides channel-level parallelism. PATTON therefore applies Case Ch first and Case C second after all token positions in a Value block have been filled. This Case Ch$\rightarrow$C order first exposes channel parallelism and then preserves a long, contiguous token reduction within each channel, minimizing WRBIAS and RDMAC transfers during SV.

\begin{table*}[t]
    \centering
    \caption{Prior PIM systems' support for efficient production LLM serving requirements.}
    \label{tab:prior-work-comparison}
    \renewcommand{\arraystretch}{1.15}
    \setlength{\tabcolsep}{5pt}
    \scriptsize
    \newcommand{\cmark}{\textcolor{green!60!black}{\checkmark}}
    \newcommand{\xmark}{\textcolor{red!75!black}{\raisebox{-0.05ex}{\scalebox{1.25}{\ensuremath{\boldsymbol{\times}}}}}}
    \begin{tabularx}{\textwidth}{l c*{5}{>{\centering\arraybackslash}X}}
        \toprule
        \textbf{Work} & \textbf{Year} & \textbf{Dynamic Memory Allocation} & \textbf{PagedAttention Support*} & \textbf{Explicit Ch--Bk--Row--Col Mapping} & \textbf{Decode Stage Single Token KV Cache Write Efficiency} & \textbf{PU Modification$^{\dagger}$} \\
        \midrule
        NeuPIMs~\cite{neupims}   & 2024 & \cmark & $\triangle$ & \xmark & \textcolor{red!75!black}{Not Considered} & \textcolor{green!60!black}{Unmodified} \\
        AttAcc~\cite{attacc}   & 2024 & \xmark & \xmark & \xmark & \textcolor{red!75!black}{Not Considered} & \textcolor{green!60!black}{Unmodified} \\
        CENT~\cite{cent}         & 2025 & \xmark & \xmark & \xmark & \textcolor{red!75!black}{Not Considered} & \textcolor{green!60!black}{Unmodified} \\
        BlockPIM~\cite{blockpim} & 2025 & \cmark & \cmark & \xmark & \textcolor{red!75!black}{Not Considered} & \textcolor{green!60!black}{Unmodified} \\
        PAISE~\cite{paise}       & 2025 & \xmark & \xmark & \cmark & \textcolor{red!75!black}{Not Considered} & \textcolor{green!60!black}{Unmodified} \\
        AttenPIM~\cite{attenpim} & 2025 & \xmark & \xmark & \cmark & \textcolor{green!60!black}{Utilized Burst} & \textcolor{red!75!black}{Modified} \\
        L3~\cite{l3}             & 2025 & \xmark & \xmark & \cmark & \textcolor{green!60!black}{Utilized Burst} & \textcolor{red!75!black}{Modified} \\
        PIMphony~\cite{pimphony} & 2026 & \cmark & \xmark & \xmark & \textcolor{red!75!black}{Not Considered} & \textcolor{green!60!black}{Unmodified} \\
        \textbf{PATTON}                     & 2026 & \cmark & \cmark & \cmark & \textcolor{green!60!black}{Efficient} & \textcolor{green!60!black}{Unmodified} \\
        \bottomrule
    \end{tabularx}
    \vspace{2pt}
    \begin{minipage}{\textwidth}
        \scriptsize\itshape
        *Full PagedAttention support includes both prefix sharing across concurrent requests and prefix caching across requests or conversation turns. $\triangle$ denotes systems that manage the KV cache in chunks or pages but bind them to individual requests. Without request-independent logical block IDs and reusable zero-reference blocks, this organization cannot support either form of prefix reuse and therefore does not provide full PagedAttention support.
        \par
        $^{\dagger}$PU Modification indicates whether the PIM processing unit must support functionality beyond the MAC operation provided by commodity PIM, including multiple accumulation dataflows and input casting methods not available in commodity PIM.
    \end{minipage}
\end{table*}

\textbf{KV cache write conflict.} With Case Ch$\rightarrow$C, the elements along the head dimension of a newly generated Value vector are distributed across rows, so writing the vector activates all $d_{\mathrm{head}}/16$ rows. Each activation writes only one element per bank before the row is closed. The layout therefore converts one Value write into several poorly utilized row activations. As shown in Fig.~\ref{fig:wrkvoverhead}(b), our characterization finds that KV cache write latency reaches about 30\% of the combined QK$^\top$ and SV latency with this layout. Production PIM serving must therefore optimize physical placement for both attention execution and decode stage single token KV cache writes.

Case R instead groups a Value block's parallel subblocks by row. A newly generated token's Value vector can then be written after activating only one row. During SV, however, adjacent bursts correspond to different outputs along the head dimension rather than one continuous reduction. The runtime must retrieve and reinitialize the accumulator between these bursts, while corresponding reductions from later blocks reside in different rows. Case R improves KV cache write efficiency at the cost of GEMV efficiency and additional I/O transfers.

\textit{No single static Value layout simultaneously preserves long GEMV reductions and minimizes row activations for single token writes.} We observe that single token Value writes target the block currently being filled, whereas blocks whose token positions are all filled no longer receive these writes. This separation suggests using Case R while a block is being filled and Case Ch$\rightarrow$C after it is filled, resolving the direct conflict between KV cache write efficiency and GEMV efficiency.

\textbf{Capacity utilization conflict.} Case C extends a request's SV reduction by placing subsequent Value blocks in adjacent columns of the same rows. Preserving this placement as the sequence grows raises a memory allocation question: should the runtime reserve the entire 2\,KB row in every ChPerHead channel for the request? Doing so guarantees that future Value blocks can extend the request's reduction without interruption in each channel while all ChPerHead channels execute in parallel. However, the reservation is made before decoding finishes, when the final sequence length is unknown. If the request never fills the row, the unwritten columns remain unavailable to other requests, reducing capacity efficiency and potentially evicting useful resident or prefix-cached blocks despite unused physical space. Fig.~\ref{fig:wrkvoverhead}(c) illustrates the cost incurred by prior PIM runtimes that cannot reuse cached prefixes: cache-miss prefill accounts for about 90\% of the prefill work, compared with only about 10\% for mandatory prefill. PATTON must avoid this behavior by preserving prefix caching through capacity-efficient allocation.

Alternatively, the runtime can manage the row at block granularity and allocate free columns to other requests. This improves capacity efficiency, but blocks from different requests can then interleave along the row and fragment each request's SV reduction. In our microbenchmark, dividing the same 2\,KB MAC work into eight fragments with intermediate RDMAC and WRBIAS operations increases latency by $2.18\times$ and energy by $1.34\times$ relative to one continuous 2\,KB MAC sequence. Production PIM serving must therefore use different Value layouts while a block is being filled and after it is filled. It also needs a block-sized physical unit that fixes each logical block's write addresses and attention commands, together with a coarser allocation unit that groups several blocks to preserve useful reductions without reserving an entire row.

\subsection{Limitations of Prior Works}

Table~\ref{tab:prior-work-comparison} compares prior PIM-based LLM systems against the requirements established above. Existing systems separately advance attention acceleration, dynamic memory management, or explicit physical mapping.

AttAcc~\cite{attacc} establishes a heterogeneous xPU+HBM-PIM architecture that assigns compute-intensive model operations to the xPU and memory-bound attention to PIM. It represents the KV cache as a large tensor and maps it to a large HBM-PIM memory region at initialization, limiting support for dynamic, block-level KV cache management. NeuPIMs~\cite{neupims} adds dual row buffers and runtime pipelining to overlap NPU and PIM stages, together with page-based KV cache allocation. However, it leaves the exact block-to-PIM mapping implicit, and its pages remain bound to individual requests. CENT~\cite{cent} scales PIM and PNM resources through CXL and maps complete transformer blocks across the system, but its KV cache organization reserves capacity for a fixed maximum context length rather than allocating blocks on demand.

BlockPIM supports dynamic allocation and PagedAttention by distributing fixed-size KV cache blocks across channels~\cite{blockpim}. Its channel--row mapping ties the supported block size to the hardware organization, making allocation inherently coarse-grained and leaving the placement of individual tokens within each block unspecified. PIMphony uses dynamic command scheduling and an address translation table on the PIM module to allocate KV cache capacity according to the actual token length~\cite{pimphony}. It manages the cache in request-bound, 1\,MB row-spanning chunks without clarifying how individual tokens fill each chunk.

The aforementioned works leave the block-to-PIM mapping implicit by representing the KV cache as large tensors, pages, or chunks and the PIM address space only at channel and row granularity. This abstraction does not determine (1) the exact physical layout of each logical block and token, (2) how Cases Ch, C, and R should be ordered, or (3) how the MAC and GB commands should be issued. Moreover, most were not designed around repeated multi-request block allocation and reclamation, and therefore do not examine how row ownership affects usable KV cache capacity or fragments one request's SV reduction.

Other prior works make the block-to-PIM mapping explicit, but either pre-allocate KV cache capacity for the maximum context length or require significant PU modifications. PAISE~\cite{paise} maps Key and Value within a large pre-allocated KV cache region. Within this region, it determines the placement of each token's Key and Value vectors based only on GEMV efficiency. PAISE evaluates an idealized serving workload where all batched requests are effectively identical, with the same prompt length and decoding progress. This artificially simplifies KV cache writes, enabling contiguous bulk memory copies and highly efficient burst accesses. AttenPIM~\cite{attenpim} and L3~\cite{l3} additionally make Value writes efficient by using a full burst as the minimum write unit and placing the bursts across rows or banks. Their layouts realize SV through an outer-product-based reduction, which requires processing-unit functionality beyond the MAC operation available in commodity PIM; they also lack dynamic allocation and PagedAttention support. Thus, prior works provide individual pieces of the required solution, but not one runtime that connects production block management to explicit physical layouts while jointly providing GEMV efficiency, KV cache write efficiency, and capacity efficiency on unmodified commodity PIM.

\begin{figure*}[t]
    \centering
    \includegraphics[width=\textwidth]{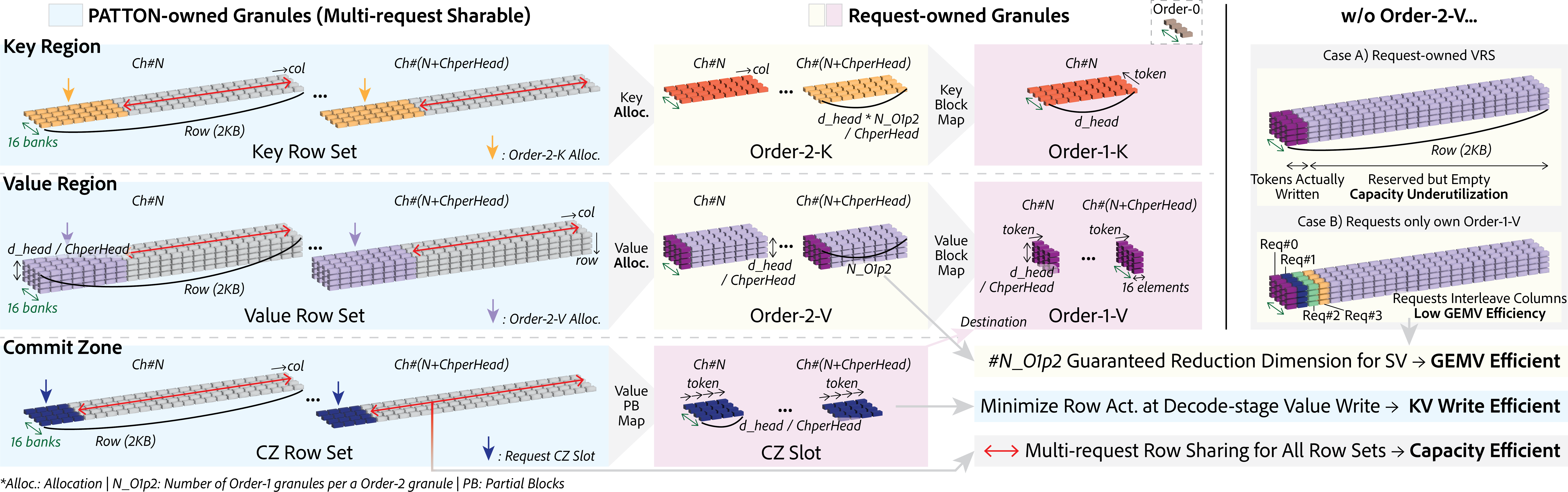}
    \caption{PATTON's hierarchical granule allocation across the KRS, VRS, and CRS. Order-1 granules bind logical blocks one-to-one to physical placements and commands; Order-2 granules group them for efficient GEMV execution and capacity use. Each CRS contains CZ slots, which stage partial Value blocks before committing them to Order-1-V.}
    \label{fig:granuleallocation}
\end{figure*}

\subsection{Design Requirements}

These observations establish four requirements for a production PIM runtime. First, the runtime must bind each logical token block to a block-sized physical unit with exact channel, bank, row, and column coordinates. This binding must determine DRAM write addresses, GB formation, and the command sequences that realize QK$^\top$ and SV. Second, because QK$^\top$ and SV have different reduction and parallel dimensions, the runtime must use different physical layouts for Key and Value and distribute their blocks across channels and columns accordingly. Third, Value placement must jointly achieve GEMV efficiency, KV cache write efficiency, and capacity efficiency: a block requires different layouts while it is being filled and after it is filled, while a coarser allocation unit must group the block-sized mappings so that multiple requests can share row capacity without frequently fragmenting SV. Finally, these mechanisms must operate through the existing PIM command set without modifying the processing units. PATTON satisfies these requirements through hierarchical granule allocation and a Commit Zone, described in the following section.

\section{PATTON}
\label{sec:patton}

\subsection{Overview}

PATTON enables production LLM serving on commodity PIM through hierarchical granule allocation: Order-0 captures the minimum valid parallel execution unit, Order-1 composes these units into block mappings, and Order-2 partitions Row Sets for request allocation. This hierarchy connects block-to-PIM mapping, physical memory allocation, and command generation. PATTON manages physical placement, while vLLM retains request scheduling and logical KV cache management, including prefix sharing and caching~\cite{vllm,vllm-github,cachedattention}.

This separation creates two synchronized paths. On the memory management path, PATTON receives logical block allocation and reclamation events and maintains the corresponding Key and Value placements. On the execution path, it resolves the logical block IDs and token offsets of the current forward pass through this metadata to generate DRAM writes and command sequences for QK$^\top$ and SV. The metadata remains valid until eviction.

Per transformer layer, PATTON divides the physical address space into Key, Value, and Commit Zone (CZ) regions, organized as \emph{Key Row Sets} (KRSs), \emph{Value Row Sets} (VRSs), and \emph{CZ Row Sets} (CRSs). KRSs and VRSs contain request-owned granules for QK$^\top$ and SV and can be shared by multiple requests. The Commit Zone contains PATTON-owned CRSs, each divided into CZ slots for efficient writes to partial Value blocks. Fig.~\ref{fig:granuleallocation} shows this hierarchy.

\subsection{Hierarchical Granule Allocation}

PATTON derives fixed, aligned granules from PIM execution constraints. Every valid placement is a known granule position, so compact bitmaps represent region availability. Allocation and reclamation select or release these positions and update ownership metadata instead of fitting irregular blocks into arbitrary channel-bank-row-column regions. 

For clarity, we explain PATTON using the representative configuration established in Section~\ref{sec:motivation}: vLLM's default 16-token block~\cite{vllm,vllm-github}, a $d_{\mathrm{head}}$ of 128 BF16 elements (eight bursts), and the AiM GDDR6 organization with 32 channels, 16 banks per channel, a 2\,KB row, and a 32\,B burst~\cite{aim}. PATTON is reconfigurable across PIM organizations and derives its granule shapes from the model and device parameters. Its only hardware dependent recommendation to vLLM is to choose a token block size at least as large as the bank count per channel and the element count per burst, enabling full bank parallelism for Key and a full burst reduction for Value. Both quantities are  below 32 in contemporary DRAMs~\cite{gddr6,lpddr5,ddr4,ddr5,hbm2}.

\paragraph{ChPerHead Channel Allocation.}

Each KV head is assigned a ChPerHead channel group. Across groups, only the data differ; memory allocation, address mapping, and command sequences are identical because KV heads execute independently in parallel. Within each group, PATTON applies the Case Ch, C, and R placement choices from Section~\ref{sec:motivation} to maximize channel parallelism and reduce row activations. Least occupied selection with round robin tie breaking balances channel use.

\paragraph{Granule Hierarchy.}

The smallest physical unit is an \emph{Order-0} granule: one burst in every bank at a channel. This is the minimum valid parallel PIM execution unit because using fewer banks sacrifices bank-level parallelism, while a reduction shorter than one burst still executes a full burst and wastes computation. An \emph{Order-1} granule is block-sized: an Order-1-K (O1K) or Order-1-V (O1V) corresponds one-to-one to a logical Key or Value block. An O1K composes Order-0 granules along one row's columns, whereas an O1V composes them at one burst column across eight Value rows distributed by ChPerHead. These bindings determine token write locations and the physical address ranges used to generate QK$^\top$ and SV commands. Each KRS and VRS is partitioned into \emph{Order-2} granules, the largest request-owned class. An Order-2-K (O2K) groups O1Ks within a KRS, whereas an Order-2-V (O2V) groups consecutive O1V columns within a VRS. Order-2 allocation packs block mappings into as few rows as possible, preserves useful GEMV spans, and allows multiple requests to share a KRS or VRS. We set $N_{\mathrm{O1p2}}=16$, which provides the best tradeoff between capacity efficiency and SV GEMV efficiency. O2K and O2V are equal in size, as are O1K, O1V, and a CZ slot. This alignment gives Key and Value the same allocation granularity and fragmentation, preventing either cache from independently triggering eviction during new allocations.

\subsection{Key Cache Mapping and Execution}

PATTON maps the tokens of a Key block to the banks of an O1K. An O1K spans 16 banks and $d_{\mathrm{head}}$ elements, while an O2K spans ChPerHead channels and 16 banks with a width of $d_{\mathrm{head}}N_{\mathrm{O1p2}}/\mathrm{ChPerHead}$ bursts. Within each bank, one Key vector occupies consecutive burst columns. A token offset selects a bank, while the O1K base column identifies the vector's first burst.

\begin{figure*}[t]
    \centering
    \includegraphics[width=\textwidth]{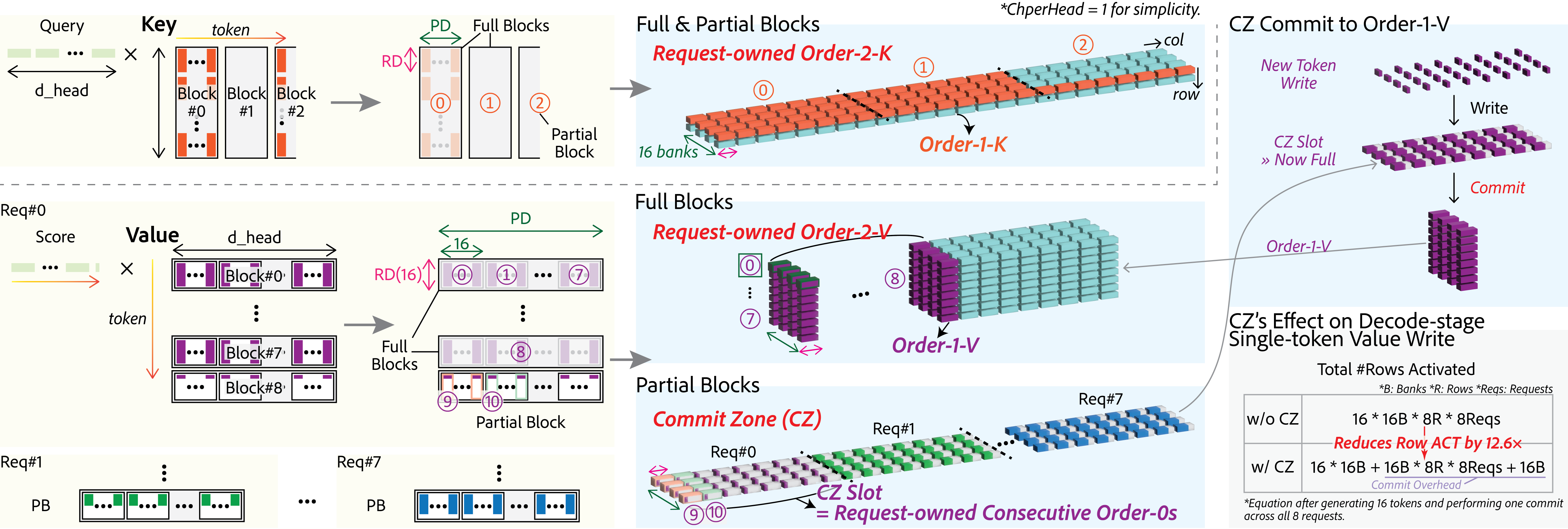}
    \caption{PATTON's mapping of logical Key and Value blocks to hierarchical granules. Full blocks map to O1Ks and O1Vs while partial Value blocks map to CZ slots until committed to O1V. The Commit Zone reduces row activations for single token Value writes by $12.6\times$, including commit overhead.}
    \label{fig:pattonmapping}
\end{figure*}

Within a KRS, PATTON distributes O1Ks inside an O2K across ChPerHead channels and then fills available columns before allocating another O2K. This implements Case Ch$\rightarrow$C while preserving each O1K's logical block binding. Fig.~\ref{fig:pattonmapping} shows consecutive Key blocks in one request-owned O2K.


For QK$^\top$, a Query of $d_{\mathrm{head}}$ elements does not fill the 2\,KB GB, so PATTON duplicates it at the column offsets of the O1Ks in a row. For each Key block, it initializes the accumulator with WRBIAS, issues MACAB commands over the head dimension, and retrieves the 16 token scores with RDMAC. Consecutive O1Ks represent independent outputs, so each block retains its own WRBIAS and RDMAC boundary.

\subsection{Value Cache Mapping and Commit Zone}

\paragraph{Value Blocks in Order-1-V and Order-2-V.}

A full 16-token Value block uses the GEMV-optimized Case Ch$\rightarrow$C layout from Section~\ref{sec:motivation}. A VRS spans ChPerHead channels and $d_{\mathrm{head}}/\mathrm{ChPerHead}$ rows. Across these channels, rows, and 16 banks, an O1V is one burst wide, while an O2V is $N_{\mathrm{O1p2}}$ bursts wide. Each row computes 16 head-dimension outputs in parallel, while each burst's 16 elements form the block's token reduction.

Each O1V fixes one logical Value block's placement and commands. PATTON partitions a VRS into independently allocatable O2Vs. Within each O2V, adjacent O1Vs extend the token reduction along each row. The Order-2 boundary therefore guarantees a contiguous group of Value blocks while leaving the remaining O2Vs available to other requests, avoiding both whole-VRS reservation and interleaving at every O1V boundary.

During SV, PATTON places the score vector in the GB at the O1V column offsets. For each row of an O2V, it initializes the accumulator, issues MACAB operations across the resident O1Vs, and retrieves 16 outputs. Consecutive O1Vs within an O2V extend the token reduction. When adjacent O2Vs in the same VRS belong to the same request, PATTON continues the MAC command sequence across the O2V boundary without intermediate result transfers; an O2V belonging to another request requires a new command sequence.


\paragraph{Partial Blocks in the Commit Zone.}

O1V is efficient only after all 16 token positions of a block are available. Directly placing a partial block in O1V would activate all eight rows for every single token Value write. PATTON instead assigns the final, partially filled Value block of each request to the Commit Zone, shown at the bottom of Fig.~\ref{fig:pattonmapping}.

For a partial Value block, PATTON reserves its final O1V destination and a free CZ slot under the same ChPerHead allocation. A CZ slot spans ChPerHead channels and 16 banks with a width of $d_{\mathrm{head}}/\mathrm{ChPerHead}$ bursts. The reserved O1V preserves the order of consecutive block columns required by SV, while the CZ slot remains the visible placement until the commit completes.

Each single token write stores one element per bank in every Order-0, producing a striped address pattern. In each channel of the ChPerHead group, the column address for token offset $t$ and local Order-0 index $i$ is
\begin{equation}
    c_{\mathrm{CZ}}(i,t) = c_{\mathrm{slot}} + 16i + t,
    \qquad 0 \leq i < \frac{d_{\mathrm{head}}}{\mathrm{ChPerHead}}.
\end{equation}
Because these positions share one row within each channel, PATTON writes them under one row activation per channel. PATTON therefore dedicates each CZ slot to one partial logical Value block. Packing multiple blocks into a slot would reduce the number of requests sharing its row, increase row activations for decode stage Value writes, and require a larger Commit Zone at the expense of Key and Value capacity.

CRSs are shared across requests, with allocation performed at the granularity of a CZ slot. PATTON fills existing CRS rows before reserving new ones and groups writes by row to reuse activations. Partial Value blocks participate in decode stage SV directly from their CZ slots, with each Order-0 initialized, executed, and retrieved separately.

\paragraph{Committing a Filled CZ Slot.}

After token offset 15 is written, PATTON \emph{commits} the CZ slot to its reserved O1V destination by writing its Order-0 granules across the eight destination rows. It publishes the O1V placement after the transfer completes and returns the CZ slot to the free pool.

A commit occurs only after every 16 decode steps, when the current Value block becomes full, whereas a single token Value write occurs at every decode step. The Commit Zone optimizes this frequent operation by staging the partial Value block in a layout that substantially reduces the number of activated rows. Committing the full block to O1V introduces additional row activations, but this cost is amortized over the preceding 16 writes. As shown in Fig.~\ref{fig:pattonmapping}, the Commit Zone reduces the total row activations for single token Value writes by $12.6\times$, even when the periodic commit overhead is included.

\subsection{Mapping Metadata and Block Lifecycle}

PATTON indexes placement metadata by vLLM logical block ID. Each entry records the O1K placement, the visible O1V or CZ slot placement, and any O1V destination reserved for a block currently in a CZ slot. Per-request state tracks referenced blocks and partially occupied O2Ks and O2Vs. The metadata determines \emph{where} a block resides; the token offset supplied by vLLM determines the position written within that block.

During prefill, PATTON binds previously unseen logical blocks and writes the token positions scheduled by vLLM. During decode, it allocates space only when vLLM appends a new logical block. Other single token writes target existing placements.

Prefix sharing and caching reuse the same physical placement. Sharing increments a full block's reference count instead of allocating or copying it; caching keeps a zero-reference full block mapped for later reuse until eviction. Partial Value blocks in CZ slots remain private and writable. Eviction unbinds O1K and O1V, returning an O2K or O2V to the free pool after its final Order-1 granule is removed. Terminating a request with a partial block in a CZ slot releases both the slot and its reserved O1V destination.


\section{Evaluation}
\begin{figure*}[t]
    \centering
    \includegraphics[width=\textwidth]{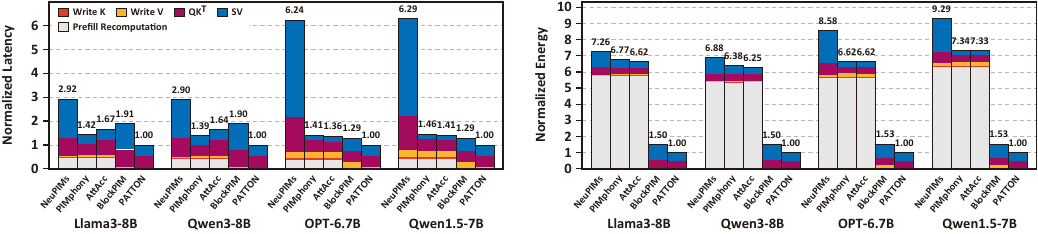}
    \caption{Latency (left) and energy (right) of attention and runtime-induced prefill recomputation across Llama3-8B, Qwen3-8B, OPT-6.7B, and Qwen1.5-7B, normalized to PATTON.}
    \label{fig:evaluation-all-breakdown}
\end{figure*}
\begin{figure}[t]
    \centering
    \includegraphics[width=\columnwidth]{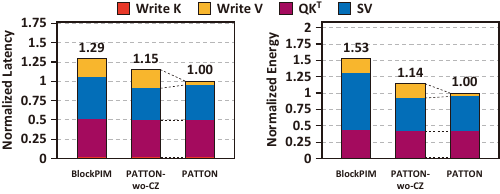}
    \caption{Attention latency (left) and energy (right)
of BlockPIM, PATTON-wo-CZ, and PATTON on Qwen1.5-7B,
normalized to PATTON.}
    \label{fig:evaluation-Commit-Zone}
\end{figure}
\begin{figure}[t]
    \centering
    \includegraphics[width=\columnwidth]{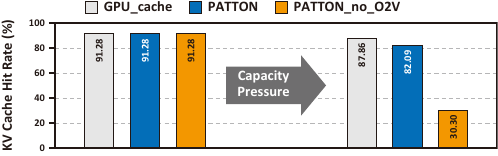}
    \caption{KV cache hit rate of GPU\_cache, PATTON, and PATTON\_no\_O2V under normal and capacity-pressure conditions.}
    \label{fig:evaluation-cache-hit-rate}
\end{figure}

\subsection{Dataset and Models}
\label{sec:method-dataset-model}

The evaluation uses ShareGPT~\cite{sharegpt}, a collection of real conversations with diverse input and output lengths. Each conversation is served through vLLM as a multi-turn chat, with conversation history accumulating across successive turns.
A target concurrency of 8 and a live request pool of 16 are used for OPT-6.7B and Qwen1.5-7B; the corresponding values are 32 and 64 for Qwen3-8B and Llama3-8B~\cite{zhang2022opt,yang2025qwen3,grattafiori2024llama,bai2023qwen}. Requests in the live pool can enter the active
batch whenever the scheduler has available capacity, producing sustained arrivals and departures rather than a fixed batch. Each configuration is run for more than 10,000 vLLM scheduler steps, and mean values are reported over the run. This duration captures long-term request and cache behavior while keeping the simulation time practical.
AttAcc and NeuPIMs intrinsically tie the maximum context length to the hardware configuration, which limits it to 2,048 tokens in the evaluated setup. This limit is therefore applied when comparing PATTON with these baselines. PATTON itself imposes no hardware-defined limit on context length.

\subsection{Runtime Layouts}
\label{sec:method-evaluation-setting}
\label{sec:runtime-layouts}

Prior PIM-based LLM systems do not provide a runtime interface that translates vLLM's logical KV cache operations into device-level PIM commands. PATTON is
therefore the only evaluated runtime that natively connects vLLM's dynamic block lifecycle to the PIM device. To compare physical mappings under dynamic request arrivals, the KV cache placement and PIM kernel schedules
of NeuPIMs, BlockPIM, AttAcc, and PIMphony are implemented behind PATTON's common command generation interface~\cite{neupims,blockpim,attacc,pimphony}. Each translated runtime receives the same kernel requests and generates PIM commands using its own physical KV cache layout.

Several baselines specify how requests, blocks, or chunks are distributed but do not provide the exact channel-bank-row-column placement of each Key and Value vector. For these cases, the GEMV-optimized Key and Value layouts from Section~\ref{sec:motivation} are applied wherever they are compatible with the documented design, giving each baseline a favorable physical mapping. The defining features of each system are retained: BlockPIM assigns a block to one least-filled channel~\cite{blockpim}; NeuPIMs assigns a request to one least-loaded channel~\cite{neupims};
AttAcc uses fixed channel partitions and statically reserves per-head capacity for the maximum context length~\cite{attacc}; and PIMphony retains
Token-Centric Partitioning and request-owned DPA chunks, with channels divided among KV heads according to ChPerHead in the single-module configuration~\cite{pimphony}.

vLLM's block-level KV cache management, prefix sharing, and prefix caching are not retrofitted into prior systems that do not specify these mechanisms. Doing
so would require inventing allocation policies that these systems do not define, making the comparison unrepresentative of their documented designs. For NeuPIMs, BlockPIM, and PIMphony, which describe page-, block-, or chunk-based KV cache partitioning, the evaluation preserves
their specified allocation granularity and request ownership. In the evaluated configuration, BlockPIM's hardware-bound KV cache block size is 128 tokens, and it does not support smaller block sizes.
AttAcc retains its statically provisioned, request-owned KV cache organization.
BlockPIM and PATTON support block-level allocation and prefix caching and thus
utilize vLLM's scheduler and KV cache manager. For NeuPIMs, AttAcc, and PIMphony, a lifecycle emulator replaces these components while preserving each runtime's
native allocation semantics.

\paragraph{PIM Configuration.}
All runtimes use the same 16\,GiB AiM configuration with one module, 32 channels, 16 banks per channel, 16,384 rows per bank, 2\,KB rows, and 32\,B bursts.
The Ramulator2-based AiM simulator models DRAM command timing using its configured timing parameters~\cite{cent}. The simulator reports memory-system cycles and per-channel command counts, which are subsequently used to calculate modeled PIM energy.

\subsection{Performance and Energy Evaluation}
\label{sec:method-performance}

PATTON is evaluated against NeuPIMs, PIMphony, AttAcc, and BlockPIM across Llama3-8B, Qwen3-8B, OPT-6.7B, and Qwen1.5-7B, with results for each model normalized to PATTON. All cross-model averages are computed using the geometric mean. Because the evaluated systems assume different host xPU configurations, host-side execution unrelated to KV cache management is excluded from the comparison. The evaluation instead isolates the costs that reflect how each runtime uses PIM for KV cache storage and decode stage attention acceleration: PIM attention execution and GPU prefill recomputation caused by runtime KV cache misses.
Fig.~\ref{fig:evaluation-all-breakdown} decomposes PIM attention latency and energy into Key and Value writes, QK$^\top$, and SV. Runtime-induced prefill recomputation is measured on an NVIDIA A100 80\,GB PCIe GPU~\cite{nvidia_a100_80gb}. Its incremental latency and energy are isolated using two measurements. The first measures the complete prefill workload, including both the prefill required even with vLLM's native GPU KV cache and the additional recomputation caused by runtime KV cache misses. The second measures only the required prefill. Subtracting the second result from the first accounts for the nonlinear relationship between prefill cost and token count.

\paragraph{Performance Comparison with the Baselines.}
Fig.~\ref{fig:evaluation-all-breakdown} reports the normalized latency and energy of PATTON and the translated runtimes. PATTON delivers average speedups of $4.27\times$, $1.42\times$, $1.52\times$, and $1.57\times$ over NeuPIMs, PIMphony, AttAcc, and BlockPIM, respectively, together with energy-efficiency improvements of $7.94\times$, $6.77\times$, $6.70\times$, and $1.51\times$. PATTON achieves high GEMV efficiency through minimal row activations, contiguous SV reductions, and channel-parallel mapping, while being the only evaluated design to optimize decode stage KV writes. More specifically, the sources of these gains differ across baselines. Compared with NeuPIMs, PIMphony, and AttAcc, PATTON's fine-grained physical allocation and block-level cache management together eliminate nearly all runtime-induced prefill recomputation. Against NeuPIMs, PATTON also avoids request-level channel confinement by distributing KV blocks across channels.

BlockPIM, like PATTON, avoids runtime-induced prefill recomputation through block-level KV cache management. However, its least-occupied-channel policy considers only the occupancy at allocation time, providing no guarantee that subsequent allocations and reclamations will preserve channel parallelism. PATTON improves on BlockPIM through more efficient KV writes, sustained channel parallelism under dynamic request activity, and contiguous Value placement that enables continuous SV reduction across blocks.

\subsection{Ablation Studies}
\label{sec:ablation-studies}

\paragraph{Effectiveness of the Commit Zone.}
Fig.~\ref{fig:evaluation-Commit-Zone} quantifies the effectiveness of the Commit Zone using PATTON-wo-CZ and BlockPIM as references. PATTON-wo-CZ uses PATTON's final Key and Value layouts but writes partial Value blocks directly to O1V, isolating the effect of the Commit Zone. Relative to PATTON-wo-CZ, PATTON reduces overall attention latency and energy by 13.0\% and 12.4\%, respectively, while accelerating Value writes by $5\times$. Compared with BlockPIM, PATTON reduces latency and energy by
22.7\% and 34.5\%, respectively. These results demonstrate that the Commit Zone substantially reduces the cost of decode stage Value writes.

\paragraph{O2V Allocation: Balancing GEMV and Capacity Efficiency.}

PATTON without O2V further improves SV efficiency by assigning an entire VRS to each request and maximizing the uninterrupted reduction span without regard to capacity sharing. \textit{Consequently, PATTON without O2V achieves the highest raw GEMV efficiency in the evaluation.} However, whole-VRS allocation strands unused columns and sharply reduces the KV cache hit rate under capacity pressure, causing additional GPU prefill recomputation.

Fig.~\ref{fig:evaluation-cache-hit-rate} reports KV cache hit rates under normal load and capacity pressure with more than 250,000 live context tokens. Without capacity pressure, all configurations achieve a 91.28\% hit rate. Under pressure, the GPU reference and PATTON retain 87.86\% and 82.09\%, respectively, whereas PATTON without O2V falls to 30.30\%. Relative to PATTON, this drop is comparable to reducing the usable cache capacity of the 16\,GiB PIM device to less than 6\,GiB. Such a loss directly opposes the objective of production serving engines, which use fine-grained block management to maximize cache utilization and retain reusable prefixes. O2V avoids this outcome while preserving efficient contiguous SV execution.
\section{Conclusion}

We presented PATTON, a runtime that translates production serving engines'
block-level KV cache operations into explicit commodity PIM placements and
commands. PATTON's hierarchical granule allocation uses Order-1 granules to
bind logical token blocks one-to-one to physical placements and commands, while
Order-2 granules group these block mappings for efficient GEMV execution and
capacity use. The Commit Zone
improves decode stage KV cache write efficiency.
Together, these mechanisms provide GEMV efficiency, KV cache write
efficiency, and capacity efficiency while supporting dynamic allocation,
prefix sharing, prefix caching, reclamation, and attention execution without
modifying the PIM processing units. Overall, PATTON accelerates the combined
attention and runtime-induced prefill workload by $1.95\times$ and improves
energy efficiency by $4.83\times$ on average relative to the evaluated
baselines.

\bibliographystyle{ACM-Reference-Format}
\bibliography{refs}

\end{document}